\documentclass[letterpaper,10pt]{article}
\usepackage{osameet2}

\usepackage{amsmath,amssymb}
\def\CN2{\mbox{$C_N^2 \ $}}
\usepackage[pdftex,colorlinks=true,bookmarks=false,citecolor=blue,urlcolor=blue]{hyperref} 

\begin{document}

\title{Optical turbulence forecast in the Adaptive Optics realm}

\author{Elena Masciadri, Alessio Turchi, Luca Fini}
\address{INAF - Osservatorio Astrofisico di Arcetri \\ Largo Enrico Fermi 5, 50125, Firenze, Italy}
\email{e-mail: elena.masciadri@inaf.it}

\begin{abstract}
(35-words maximum) In this talk I present the scientific drivers related to the optical turbulence forecast applied to the ground-based astronomy supported by Adaptive Optics, the state of the art of the achieved results and the most relevant challenges for future progresses.
\end{abstract}

\ocis{010.1330, 110.1080, 280.7060}

\section{Introduction}

The optical turbulence (OT) forecasts is an extremely challenging and crucial goal in the context of the ground-based astronomy of current top-class telescopes (8m-10m) as well as of new generation telescopes, the Extremely Large Telescopes (ELTs, order of 30m-40m), particularly those supported by adaptive optics and/or interferometry in the visible and near-infrared wavelengths. 
Adaptive Optics applied to ground-based astronomy is indeed able to provide, at present, images with angular resolution that is up to three time that achievable with Hubble on space. A Strehl Ratio as high as 80-90 \% in H band has been obtained on sky by extreme adaptive optics (XAO) systems mounted on 8-meters class telescopes (FLAO on Large Binocular Telescope \cite{esposito2010}, SPHERE on the Very Large Telescope \cite{fusco2014}, GPI on Gemini South \cite{mcintosh2014}). Equivalent or even better performance are foreseen for the Extremely Large Telescope (diameter of the order of 40~m) and this will open to scenarios of observations never done before with potential exciting discoveries in different fields of the astronomy. However, adaptive optics is strongly dependent on the optical turbulence conditions. It follows that the exploitation of the potentialities of these facilities strongly depends on our ability in forecasting, in advance with respect to the observing time, the most suitable atmospheric turbulence conditions for different scientific programs and different adaptive optics systems having different constraints with respect to turbulence. Astronomical observations related to the most challenging scientific programs require frequently excellent turbulent conditions. It is therefore mandatory to identify these temporal windows in which we can achieve the best scientific results to exploit the ELTs \cite{masciadri2013}.

\section{Optical turbulence forecast} 

Optical turbulence forecast obtained with non-hydrostatic mesoscale numerical atmospherical models represents the leading-edge frontier of the research in this field. These models allow the description of the optical turbulence in numerical terms thanks to the {\it parameterization} of the OT. This means to describe the spatio-temporal fluctuations on a miscroscopic scale, as a function of the gradient of the same parameters but space averaged over a larger (macroscopic) spatial scale, i.e. the model unit cell. In this talk I will review the most relevant scientific drivers for such a kind of studies, the most important milestones obtained during the last decades from the pioneering studies \cite{bougeault1995,masciadri1999} up to the most recent and relevant achievements \cite{hagelin2011,masciadri2017} obtained with the Astro-Meso-Nh code that proves the maturity of this technics. I will present the principle of the model calibration that have been proposed for the first time by Masciadri \& Jabouille (2001) \cite{masciadri2001} to improve the model performances and it has been used in several different variants later on in different contexts  \cite{masciadri2004,cherubini2011,masciadri2017} and the perspective for an universal calibration. I will finally present the new frontier and challenges for the optical turbulence forecast, particularly in the framework of the most sophisticated adaptive optics systems such as the wide field adaptive optics (WFAO). 

\begin{figure}[htbp]
  \centering
    \includegraphics[width=14cm]{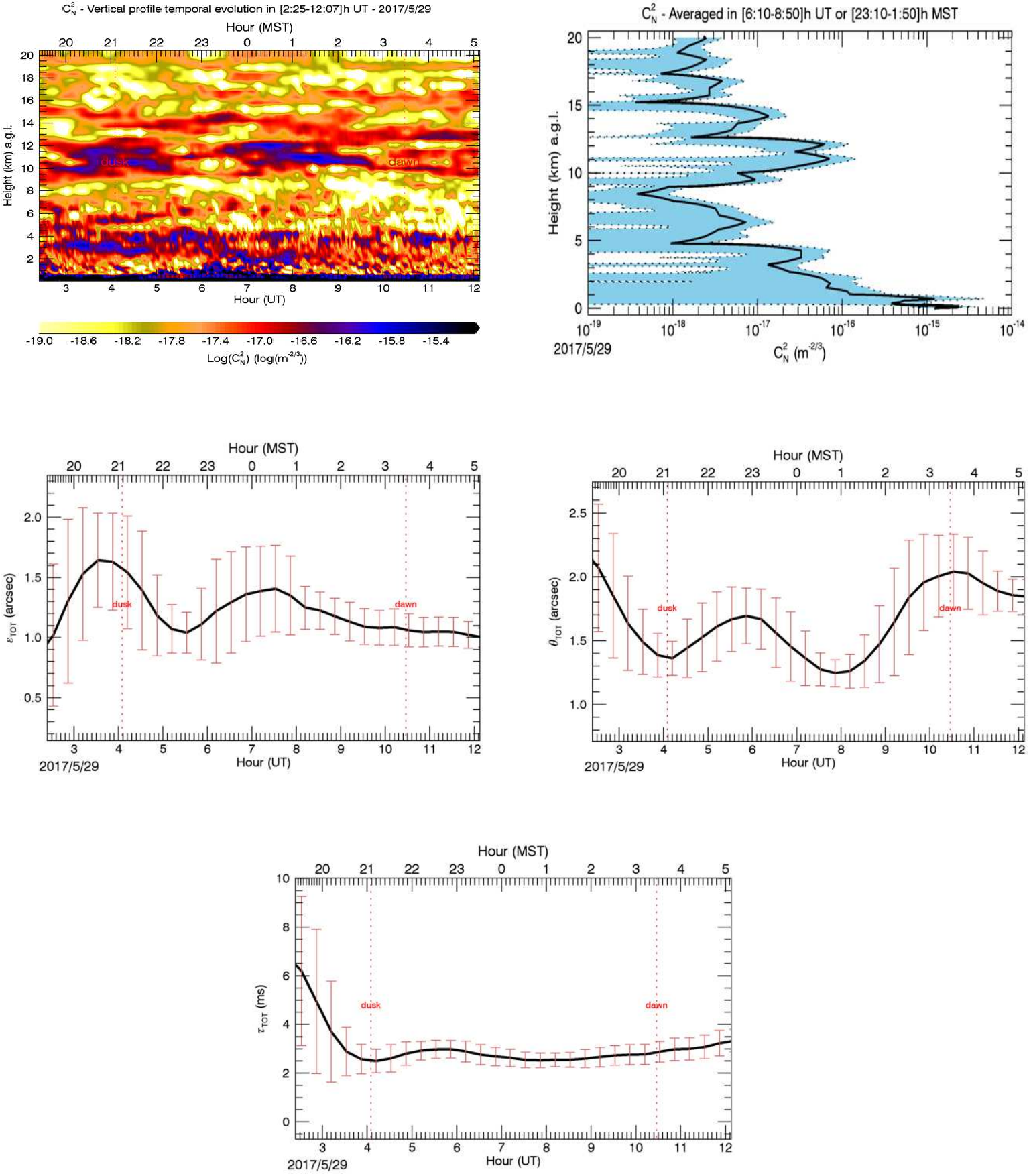}
\caption{Extracted from the ALTA Center (see text). Top raw-left: temporal evolution of the $\CN2$ profile during a night between the sunset and sunrise. Top raw-right: $\CN2$ profile averaged in a partial interval of time of the night [6:10-8:50] hours UT. The dotted lines represent the $\pm$ $\sigma$. Central raw-left: temporal evolution of the seeing related to the whole atmosphere $\varepsilon_{TOT}$ and obtained from the $\CN2$. Central raw-right: temporal evolution of the associated isoplanatic angle $\theta_{0,TOT}$ calculated on the whole atmosphere and obtained from the $\CN2$. Bottom raw: temporal evolution of the associated isoplanatic angle $\tau_{0,TOT}$ calculated on the whole atmosphere and obtained from the $\CN2$ and the wind speed\label{fig1}.}
\end{figure}

\section{Conclusions and Perspectives} 

Most relevant scientific results achieved so far indicate that it is possible, at present, to forecast the optical turbulence in application to the ground-based astronomy with reasonable good performances, certainly useful for a flexible scheduling of scientific programs and instrumentation. Automatic systems conceived to forecast the OT have already been implemented in proximity of top-class telescopes such as, for example, the ALTA Center Project\footnote{\href{http://alta.arcetri.inaf.it}{http://alta.arcetri.inaf.it}}\cite{masciadri1999,masciadri2017}, a tool expressly developed to support, in operational configuration, the scheduling of the Large Binocular Telescope (LBT) observations above Mt.Graham (Arizona) \cite{veillet2016}. This achievement is particularly exciting considering that LBT is the first precursor of the Extremely Large Telescopes. Figure \ref{fig1} shows an example of the forecasts of a few astroclimatic parameters obtained with the Astro-Meso-Nh code in the context of the flexible scheduling at the LBT Observatory through the ALTA Center \cite{masciadri2017}. The optical turbulence forecast remains a very challenging and crucial task related, from one side, to the level of accuracy and precision required for applications to the ground-based astronomy in which the dynamic range that distinguishes a good from a bad 'seeing'\footnote{The term {\it 'seeing'} is used in astronomical context to indicate the typical size of the FWHM obtained by a perfect wavefront passing through the atmosphere and focusing on a detector on the ground.} is relatively narrow (less than 1 arcsec) and, from the other side, to the level of difficulty in predicting phenomena such as the turbulence that fluctuates on spatial and temporal scales that are definitely smaller than the typical grids of the mesoscale numerical models. Results obtained so far are certainly encouraging and tell us that the right direction has been taken.

\section*{Acknowledgements}
Part of simulations performed by our group with the Meso-Nh and Astro-Meso-Nh codes have been run on the HPC facilities of the European Centre for Medium Range Weather Forecasts (ECMWF) - Project SPITFOT. We acknowledge the Meso-Nh User Supporter Team. We acknowledge the LBTO Director, C. Veillet and the LBT Board for supporting us in the development of the ALTA Center Project. We acknowledge the ESO Board of the MOSE project (P.Y. Madec, M. Sarazin, F. Kerber and H. Kuntschner) as well as J. Milli and VLT Science Operation team for their support in our studies.

\end{document}